\documentclass[prb,twocolumn,showpacs,amssymb]{revtex4}
\usepackage{graphicx}% Include figure files
\usepackage{dcolumn}% Align table columns on decimal point
\usepackage{bm}% bold math

\newcommand{\br}{{\bf r}}
\newcommand{\bB}{{\bf B}}
\newcommand{\bA}{{\bf A}}

\newcommand{\<}{{\langle}}
\renewcommand{\>}{{\rangle}}
\begin{document}

\title{Zero Temperature Glass Transition 
in the Two-Dimensional Gauge Glass Model}

\author{Marios Nikolaou and Mats Wallin}

\affiliation{Condensed Matter Theory, Department of Physics,
KTH, SE-106 91 Stockholm, Sweden}

\date{\today}
\begin{abstract}
We investigate dynamic scaling properties of the two-dimensional gauge
glass model for the vortex glass phase in superconductors with
quenched disorder.  From extensive Monte Carlo simulations we obtain
static and dynamic finite size scaling behavior, where the static
simulations use a temperature exchange method to ensure convergence at
low temperatures.  Both static and dynamic scaling of Monte Carlo data
is consistent with a glass transition at zero temperature, with
correlation length exponent given by $1/\nu=0.36 \pm 0.03$.  We study
a dynamic correlation function for the superconducting order
parameter, as well as the phase slip resistance.  From the scaling of
these two functions, we find evidence for two distinct diverging
correlation times at the zero temperature glass transition.  The
longer of these time scales is associated with phase slip fluctuations
across the system that lead to finite resistance at any finite
temperature.  The shorter time scale can be described by the form
$\tau \sim \xi^z$, with a dynamic exponent $z=2.7 \pm 0.2$, and
corresponds to local phase fluctuations.
\end{abstract}
 % (Num. Simulations)

\maketitle

\section{Introduction}
\label{sec.introduction}

Despite intense studies of the superconducting glass transition in
vortex systems, the precise nature of the transition remains
controversial.  The presence of quenched disorder in a strongly type
II superconductor in an applied magnetic field may create a
superconducting vortex glass phase with zero linear resistance, where
the vortices are localized in random positions given by the disorder
and interactions \cite{fisher89}.  Considerable theoretical and
simulation efforts have been focused on whether a superconducting
glass state exists at finite temperature, or alternatively, only at
zero temperature.  In particular the situation in two dimensions still
remains debated.  In this paper we present results from Monte Carlo
simulation and finite size scaling analysis for the glass transition
in two dimensions.  We focus mainly on the dynamic scaling properties,
which have not been studied in detail before, using a dynamic scaling
ansatz for a zero temperature transition.  From our static and dynamic
scaling analysis, we find evidence for a zero temperature glass
transition, and present various predictions for the physical
properties at the transitions.

The gauge glass (GG) model is an XY model with a random quenched
vector potential on the links of the lattice; see Eq.\ (\ref{H})
below.  The GG model contains important aspects of order parameter
symmetry, interactions, and quenched disorder, and has been frequently
studied as a model for a vortex glass transition.  In a real glass,
the system may fall out of equilibrium at the transition, but here we
focus on the equilibrium critical behavior of the model.  This enables
the study of universal scaling laws at the transition, that depend
only on the universality class.  It has not been settled if the GG
model represents the correct universality class of the vortex glass
transition, and it is thus motivated to obtain the precise critical
behavior of the GG model.  The GG model is not realistic in a few
important aspects: disorder is included only in terms of a random
vector potential, instead of a random pinning energy for the vortex
cores, and it has zero applied magnetic field.  Alternatively, a
suitable modification of the GG model as a model for the elastic
energy of a dislocated vortex lattice is discussed in Ref.\
\onlinecite{lidmar}.  Another worry is the use of a classical model
for a zero temperature transition, where quantum effects may be
important.  However, since vortices localize in the glass phase, their
quantum fluctuations are small and can be neglected, and a classical
treatment is a reasonable starting point, also at $T=0$.  This
neglects the possibility of quantum tunneling of vortices, which may
actually be important.

Considerable simulation efforts have been invested in trying to locate
the lower critical dimension $d_c$ of the vortex glass transition,
i.e., the limiting dimensionality such that a glass phase exists at
finite temperature only for $d > d_c$.  In the 3D case, several
simulations \cite{reger,olson00,katzgraber02} show that a
vortex glass phase does exist at finite temperatures in the absence of
magnetic screening.  In 2D the situation is less clear.  It has been
shown rigorously that there can be no vortex-glass order in 2D for
$T>0$ \cite{nishimori}.  For a finite screening length it is expected
that no glass phase can exist at $T>0$ \cite{bokil}.  Akino and
Kosterlitz computed the domain wall stiffness exponent $\theta$ at
zero temperature, and obtained a negative value, $\theta=-0.36$,
indicating that $T_c=0$ \cite{akino02}.  Holme and Olsson computed the
energy barrier for vortex motion across the system and argue that
$T_c>0$ \cite{holme02}.  Katzgraber and Young recently obtained
$T_c=0$ and $\theta=-1/\nu=-0.39\pm0.03$ from finite temperature
simulations \cite{katzgraber02,katzgraber03}.  In contrast, several
other studies of static quantities obtained $T_c \approx 0.2$
\cite{holme03,choi}.  Dynamic scaling in the 2D GG model has been
studied in several papers
\cite{li,kim,choi,hyman,wengel,katzgraber04}, with conflicting
results.  It is important to consider dynamic scaling in order to
obtain information about physically measurable quantities like the
resistance.  Previous studies of zero temperature scaling of dynamic
properties considered nonlinear $I-V$ characteristics \cite{hyman},
and in addition the linear resistance was found to be roughly
consistent with a simple Arrhenius form \cite{hyman,wengel}, $R \sim
e^{-K/T}$, but no further support for zero temperature dynamic scaling
has been obtained.  In contrast, Refs.\ \onlinecite{li,kim,choi}
obtained $T_c>0$ from vortex dynamics simulations.  Hence the behavior
of the GG model remain controversial, in particular for the dynamic
properties.

In this paper we study critical scaling properties at the glass
transition for both dynamic and static quantities, assuming a zero
temperature transition.  Our simulations are motivated in part by new
methods that have recently been developed.  At low temperature the
correlation times in the simulation grow very long and efficient
convergence acceleration methods become highly desirable.  The
temperature exchange method \cite{exchange,marinari} has proven very
useful for such problems, where configurations are exchanged between
different temperatures, in order to avoid getting stuck in metastable
states.  This considerably speeds up the convergence of the
simulation, and thereby enables simulation at larger system sizes and
lower temperatures.  Our scaling results for static quantities are in
good agreement with some of the previous results in the literature for
the correlation length exponent and the domain wall energy exponent
$\theta$ \cite{hyman,wengel,katzgraber02,katzgraber03,akino02}.  For
our dynamics simulations we also get valuable input from the exchange
method, even though exchange is not directly applicable for dynamics,
by using typical equilibrium configurations from the exchange runs as
input to the dynamics runs.  This considerably extends the regime of
temperatures and system sizes accessible for the dynamics simulation.
The main new results of our work are for the dynamic finite size
scaling analysis of the linear resistance and of the dynamic
correlation function for the order parameter at the zero temperature
glass transition.  We obtain correlation times from finite size
scaling of our dynamic MC data that show novel behavior.  We find
evidence for two distinct diverging correlation times, one associated
with phase slip fluctuations, and the other associated with local
phase fluctuations that are independent of the phase slip processes.

The organization of the paper is as follows: In Sec.\ \ref{sec.model}
we introduce the GG model and describe our simulation methods.  In
Sec.\ \ref{sec.scaling} we discuss the quantities calculated in the
simulation and the finite size scaling methods.  In Sec.\
\ref{sec.static} the static results are presented, and Sec.\
\ref{sec.dynamic} contains the results for the dynamic quantities.
Section \ref{sec.discussion} contains discussion and conclusions.

\section{Model and Monte Carlo simulation details}
\label{sec.model}

The gauge glass (GG) model \cite{shih,huse,ffh} is defined by the
Hamiltonian
\begin{equation}
H=-\sum_{\<i,j\>} \cos(\theta_i-\theta_j-A_{ij})
\label{H}
\end{equation}
where $\theta_i$ is the phase of the superconducting order parameter
on site $i$ of a square lattice, and $\sum_{\<i,j\>}$ denotes
summation over all nearest-neighbor pairs in the lattice.  The vector
potential enters through $A_{ij}=\frac{2\pi}{\phi_0} \int_i^j d\br
\cdot \bA$, which is a quenched random variable on each link in the
lattice, with a uniform probability distribution in $[0,2\pi)$,
corresponding to fixed random magnetic induction $\bB=\nabla \times
\bA$ through the plaquettes of the lattice.  The partition function is
$Z= \left( \prod_j \int_0^{2\pi} d\theta_j \right) \exp(-H/T)$, where
$T$ is the temperature.  The simulation uses square lattices of size
$L \times L$ with periodic boundary conditions (PBC) in both the $x$
and $y$-directions.

We also consider the GG model with different boundary conditions,
where the system repeats periodically, but with fluctuating phase
twist variables $\Delta_x, \Delta_y$ added to the respective phase
differences across the boundary of the system.  This allows for phase
twist fluctuations across the system, and enables calculation of the
resistance of the model, as is discussed below.  In this case the
Hamiltonian is
\begin{equation}
H=-\sum_{\<i,j\>}
\cos(\theta_i-\theta_j-A_{ij}-\delta_{x_i,L}\Delta_x-\delta_{y_i,L}\Delta_y)
\label{Htwist}
\end{equation}
where $\Delta_x, \Delta_y$ are integrated over $(-\infty,\infty)$ in the
partition function.  In a static simulation, the partition function is
invariant if the the phase twist at the boundary is spread out in the
bulk as a constant extra phase difference $\Delta/L$ on each link.
However, dynamically these both prescriptions are different.  We
define the phase twist fluctuations located at the boundary in order
to maintain local dynamics.

In the Monte Carlo simulations the trial moves are attempts to assign
the phase, at a randomly chosen site, a random value from a uniform
distribution in $[0,2\pi)$.  The moves are accepted with probability
$1/(1+\exp \Delta E/T)$, where $\Delta E$ is the energy difference for
the trial move.  One sweep is defined as on average one attempt to
update each phase.  In the simulations of static quantities we use a
temperature exchange method \cite{exchange,marinari}, that considerably
enhances the convergence of the simulations at low temperature and big
system sizes.  We use ten sweeps of local MC moves, followed by one
exchange sweep.  In the exchange sweep the trial moves are attempts to
exchange configurations between neighboring temperatures, where up to
39 closely spaced temperatures, with $T_n=T_{n-1}(1-1/L)$, are
simulated in parallel, sharing the same disorder realization.  The
exchange moves are accepted with probability
$\exp[-(\beta_n-\beta_{n+1})(E_{n+1}-E_n)]$, where $\beta=1/T$.  To
reach equilibrium we discard up to $6 \cdot 10^4$ MC sweeps, followed
by equally many sweeps where data is collected.  The results are
averaged over up to $10^4$ realizations of the random vector potential
$A_{ij}$.  To avoid systematic errors in the calculation of squares of
expectation values, two replicas of the system with the same disorder
are used.  We use various tests to verify that sufficient
equilibration is achieved, which will be presented in the results
section below.  In this paper we denote thermal averages by $\< \cdots
\>$, and disorder averages by $[ \cdots ]$.

In the simulation of dynamic quantities the exchange MC method does
not apply directly, since we must use a method that respects the local
nature of the dynamics of the system.  Instead we use the usual Monte
Carlo scheme with local updates of the phase.  However, also for the
dynamics runs we take advantage of the exchange algorithm in two ways.
First, we use equilibrium configurations from an exchange run as
initial configurations for the dynamics runs.  Second, during the
dynamics simulation we also recalculate the static quantities, and
compare to the same quantities from the exchange runs.  This gives an
additional test of convergence for the dynamics simulation.  We define
one time step to be one MC sweep through the system with local
updates.  In the simulation with fluctuating twists, one time step
corresponds to one sweep with local phase updates, followed by one
update of $\Delta_x$ and of $\Delta_y$.  The phase twist updates are
attempts to add a uniformly distributed random number in $(-\pi,\pi)$.
As initial configurations in the dynamics runs, we use configurations
generated in an exchange run, followed by up to $4 \cdot 10^5$ sweeps
of local MC moves where data is taken.  Occasional runs with even more
sweeps were also used to test convergence.  Disorder averages are
formed over up to $10^4$ realizations of the disorder.  We stress that
there are many dynamic universality classes, and here we only consider
that of local relaxation of the phase, which has proven useful in
previous studies of the superconducting phase transition
\cite{hyman,dynamics}.

\section{Calculated quantities and finite size scaling relations}
\label{sec.scaling}

The basic quantity in our analysis of thermodynamic properties of the
glass transition is the root mean square (RMS) current
\cite{reger,olson00}.  The current in the $x$-direction is defined as
the derivative of the free energy $F$ with respect to a phase twist
$\Delta_x$ on the boundary in the $x$-direction, for $\Delta_x=0$:
\begin{equation}
I= \left. \frac{\partial F}{\partial \Delta_x} \right|_{\Delta_x=0}=
\frac{1}{L} \sum_j \< \sin(\theta_{j+\hat{x}}-\theta_j-A_{j,j+\hat{x}}) \>
\label{I}
\end{equation}
where $j+\hat{x}$ denotes the nearest neighbor (NN) site next to $j$
in the $x$-direction.  Here the phase twist $\Delta_x$ at the boundary
has been absorbed into the bulk by adding a uniform phase twist
$A=\Delta_x/L$ on each link in the $x$-direction, which gives the
$1/L$ factor in the last equation.  The RMS current is defined as
\begin{equation}
I_{\rm rms} = [ I^a I^b ]^{1/2}
\label{Irms}
\end{equation}
where $a,b$ denote two replicas of the system sharing identical
disorder.  

In order to use the scaling properties of MC data to discriminate
between the possibilities that $T_c=0$ and $T_c>0$, we must formulate
suitable scaling assumptions for each case \cite{hyman,katzgraber02}.
At the glass transition the correlation length is assumed to diverge
as $\xi \sim |T-T_c|^{-\nu}$ if $T_c>0$, and as $\xi \sim T^{-\nu}$ if
$T_c=0$.  In the case $T_c>0$, we assume that close to $T_c$ the RMS
current obeys the finite size scaling relation \cite{olson00}
\begin{equation}
I_{\rm rms}(T,L) = \tilde{I}(L^{1/\nu}(T-T_c))
\label{Iscale0}
\end{equation}
where $\tilde{I}$ is a universal scaling function.  This scaling form
implies that the RMS current is independent of system size $L$ at
$T=T_c$, which gives a convenient criterion for locating $T_c$.  In
contrast, if $T_c=0$, the scaling relation becomes modified, since for
$T_c=0$ we have $\xi \sim T^{-\nu}=\beta^{\nu}$, and hence $\beta \sim
\xi^{1/\nu}$, which must be included in the scaling relation.  This
gives $\beta I = \partial \beta F/\partial \Delta \sim \xi^0$.  The
scaling ansatz for the RMS current valid for the case $T_c=0$ becomes
\begin{equation}
I_{\rm rms}(T,L) = L^{-1/\nu} \tilde{I}(L^{1/\nu}T)
\label{Iscale}
\end{equation}
Below we will compare both these scaling forms to our data, and find
that the scaling form with $T_c>0$ is not fulfilled by our MC data,
while the scaling form with $T_c=0$ works well.

Next we discuss the quantities that we calculate in the dynamic
simulation, and their respective scaling relations.  The aim of our
analysis is to obtain scaling relations from a dynamic scaling ansatz
that assumes $T_c=0$, and here we will only formulate the scaling
relations for this case.  The scaling relations will first be
formulated in terms of the correlation time, $\tau$, whose scaling
will in turn be discussed below.

The resistance can be calculated by using fluctuating twist boundary
conditions instead of PBC.  It is obtained from the Kubo formula for
the voltage-voltage correlation function.  We define a time dependent
resistance function as 
\begin{equation}
R(t,T,L)=  \frac{1}{2T} \sum_{t'=-t}^t \Delta t' 
\left[ \< V(t')V(0) \> \right]
\label{R}
\end{equation}
where the voltage at MC timestep $t$ is given by the Josephson
relation $V(t) = \frac{\hbar}{2e} \frac{d\Delta_x}{dt}$ (we henceforth
assume units such that $\hbar/2e=1$).  Here $d\Delta_x/dt$ is the rate
of change of the phase shift across the system in the $x$-direction,
which gives the resistance in the $x$-direction.  The usual Kubo
formula \cite{kubo} for the resistance is obtained from this function
in the limit when the summation over MC timesteps $t \to \infty$.  In
practice this limit means that the summation time has to be long
enough that the resistance is independent of $t$.  However, since this
time is going to be very long in the simulation, it is quite useful to
consider $R$ as a function of $t$ and study the scaling properties as
function of $t$.  Since the voltage scales as $V \sim 1/\tau$, we make
the finite size scaling ansatz
\begin{equation}
R(t,T,L)=\frac{1}{\tau T} \tilde{R} (L^{1/\nu}T,t/\tau)
\label{Rscale}
\end{equation}
where $\tilde{R}$ is a scaling function.

Another useful quantity is the dynamic correlation function for the
real part of the superconducting order parameter, $\Psi'=\sum_j \cos
\theta_j$, which is gauge dependent and hence not directly observable
in a superconductor, but gives a convenient measure of the dynamic
correlations of the local order in the system.  We define the time
dependent summed autocorrelation function
\begin{equation}
G(t,T,L) = \sum_{t'=-t}^t \Delta t'
\left[ \frac{\<\Psi'(t')\Psi'(0)\>}{\<\Psi'^2\>} \right]
\label{G}
\end{equation}
For $t \to \infty$, $G$ measures the correlation time $\tau$ of the
fluctuations in $\Psi'$, and we have
\begin{equation}
G(t,T,L) = \tau \tilde{G}(L^{1/\nu}T,t/\tau)
\label{Gscale}
\end{equation}

Note that all the (unknown) dynamic scaling functions contain two
arguments, which complicates their analysis.  We will make use of two
possible ways to obtain a function of one variable only, and this
function can then be conveniently analyzed.  One way is to obtain data
for $t \gg \tau$, and the second way is to fix the first argument to a
constant value by using different temperatures for each system size so
that $L^{1/\nu}T=const$ \cite{hyman}, which then permits scaling
analysis of the entire $t/\tau$ dependence.

In order to quantitatively fit the dynamic scaling formulas to MC
data, we need an ansatz for the form of $\tau$.  This is related to
the divergence of the free energy barriers at the glass transition.
The thermodynamic free energy barrier exponent $\theta$ is defined
from the interfacial free energy RMS difference, $U_L$, for changing the
boundary condition in one direction from PBC to anti-PBC.  For a
system of size $L$ we have $U_L =K L^\theta$, where $K$ is a constant.
Now, a change in free energy from a change in the boundary conditions
must scale in the same way as the current in Eqs.\ (\ref{I}) and
(\ref{Iscale}), which gives $\theta=-1/\nu$.  Thus $\theta$ can be
obtained from analysis of $I$ at finite temperatures.

A dynamic barrier exponent $\psi$ can be defined by $V_L =K L^\psi$,
which is the free energy cost to move a vortex a distance $L$ using
only local moves.  In general $V_L$ and $U_L$ do not need to coincide,
and the dynamic barrier for vortex motion across the system can be
very big.  We assume that the correlation time scales in an activated
form given by
\begin{equation}
\tau \sim \exp ( V_L/T) \sim \exp ( C/T^{1+\nu\psi})
\label{tau}
\end{equation}
where $C$ is a constant.  If $\psi=0$ this reduces to a simple
Arrhenius form, $\tau \sim \exp C/T$.  We also consider a few
other possibilities.  The usual finite temperature form of the
critical slowing down is $\tau \sim L^z$.  Another possible form is a
logarithmic barrier, $\tau \sim \exp K (\ln L) /T$.

\section{Results for static quantities}
\label{sec.static}

\begin{figure}
\includegraphics[width=8cm]{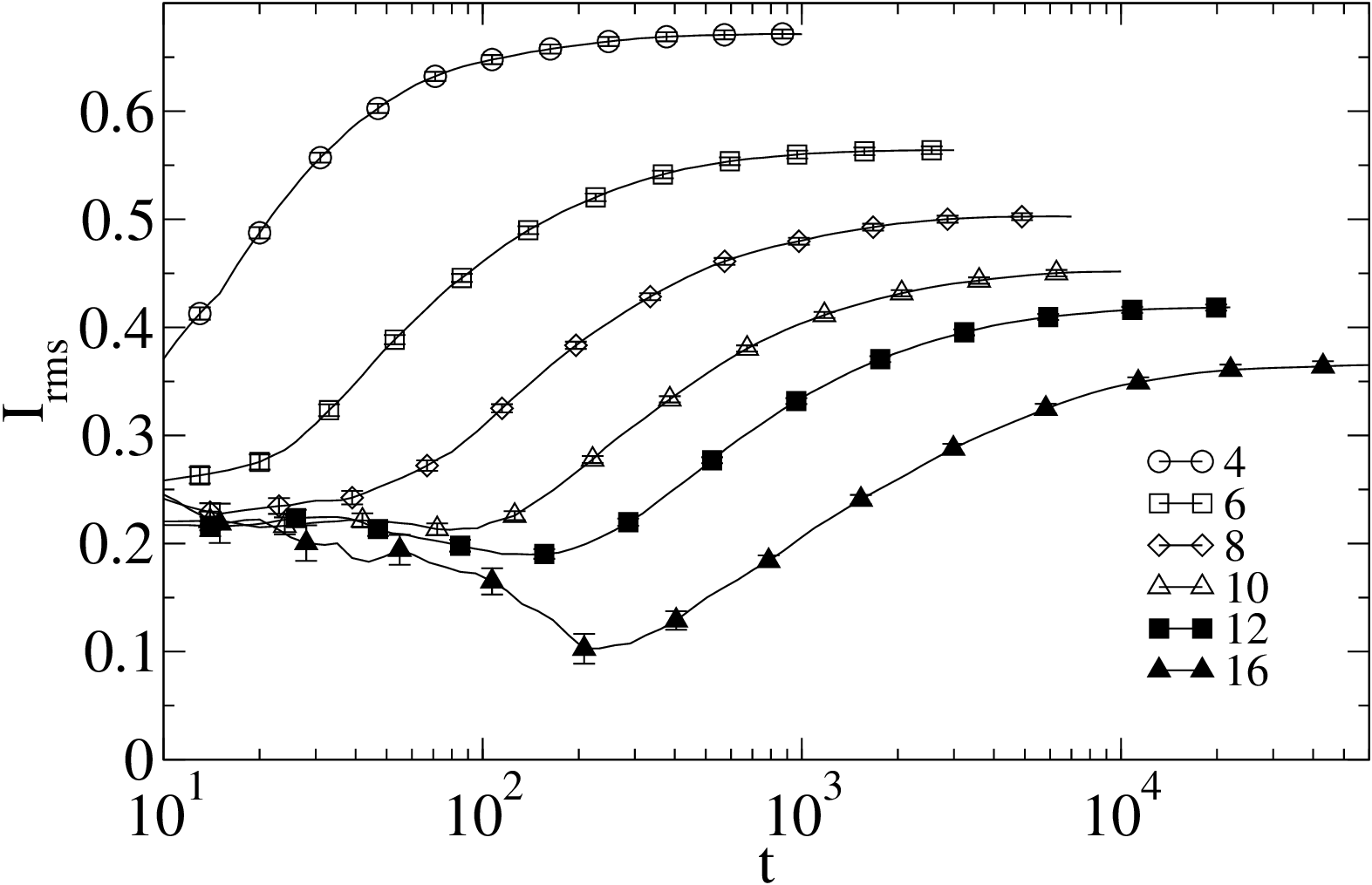}
\caption{Test of equilibration of MC data for the RMS current at $T=0.1$.
Here $t$ sweeps are taken without any measurements, followed by another $t$
sweeps where data is collected.  The saturation time in the plot gives
an estimate of the necessary equilibration times for the different
system sizes.  }
\label{fig.warmup}
\end{figure}

We first study the approach to equilibration of the exchange MC
simulation.  This is necessary in order to avoid systematic errors due
to insufficient equilibration.  We consider the RMS current, $I_{\rm
rms}$, defined in Eq.\ (\ref{Irms}).  Figure \ref{fig.warmup} shows
some of our data for the RMS current as a function of simulation
warmup time.  The data points in the figure are calculated by first
discarding $t$ MC sweeps, and then forming averages over equally many
sweeps.  In Fig.\ \ref{fig.warmup} we see that the equilibration time,
where the curves become independent of $t$, becomes very long already
for moderate system sizes $L$, despite the considerable convergence
speed-up due to the exchange method.

\begin{figure}
\includegraphics[width=8cm]{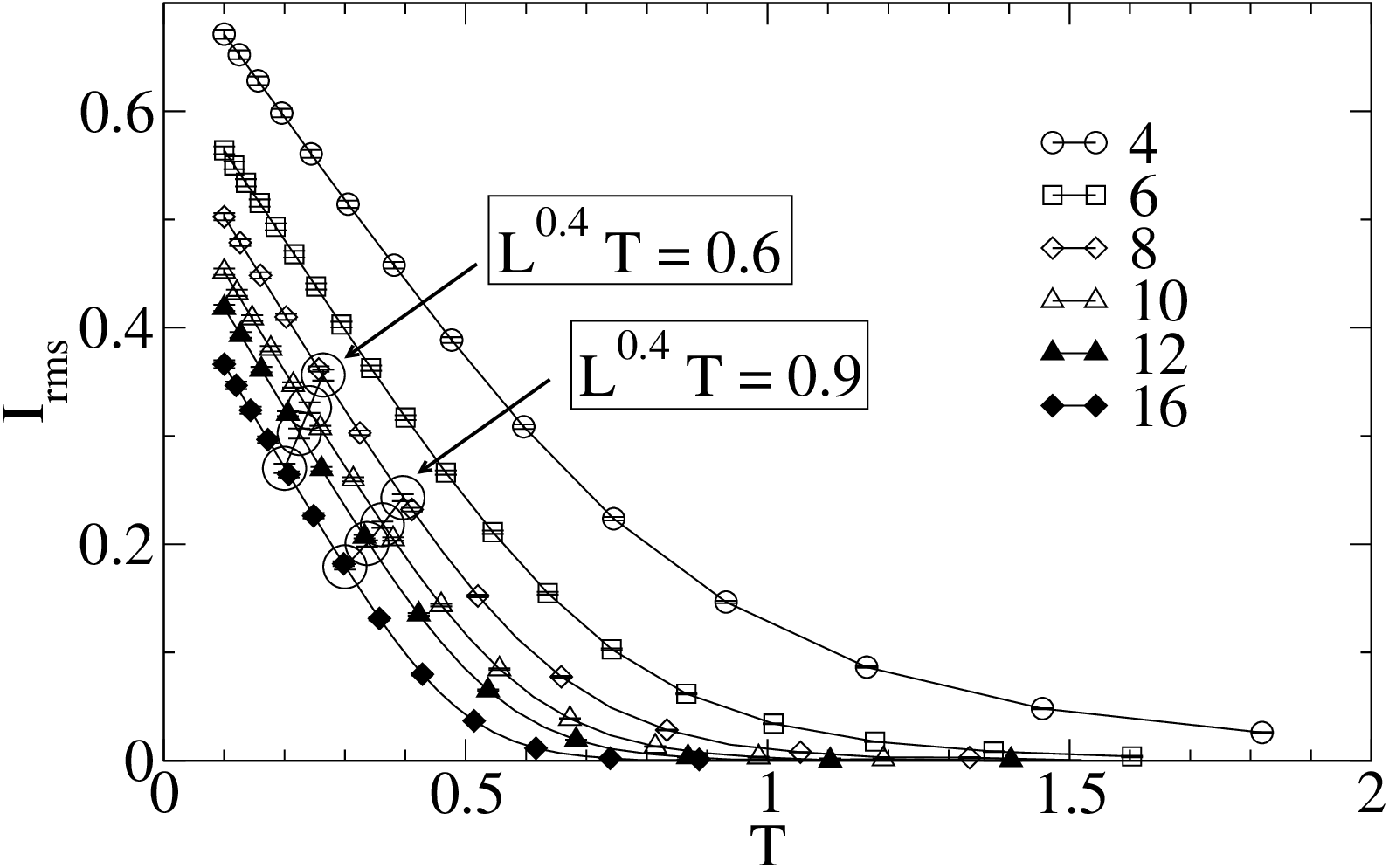}
\caption{MC data for the RMS current vs.\ temperature for different
system sizes $L$.  Finite size scaling arguments valid for $T_c >0$
(see text) show that data curves for different system sizes become
independent of $L$ at $T=T_c$, and hence intersect.  The absence of
any intersection in the plot implies that $T_c \ll 0.1$.  The data
points connected by lines are produced by the temperature exchange
method.  The circles mark data points obtained without the exchange
method, to allow calculation of dynamic quantities, and are in good
agreement with the data points obtained from the exchange method,
indicating that ensemble averages and dynamic averages tend to
coincide.  }
\label{fig.IrmsT}
\end{figure}

Next we consider finite size scaling properties of the RMS current.
Figure \ref{fig.IrmsT} shows MC data for the RMS current vs.\
temperature.  According to Eq.\ (\ref{Iscale0}) data curves for
$I_{\rm rms}$ should become system size independent and thus intersect
at $T=T_c$, if $T_c>0$.  The absence of any such intersection in the
figure indicates that $T_c$ is much smaller than the temperatures
where we have data, i.e., $T_c \ll 0.1$.  Henceforth we will therefore
assume that $T_c=0$.

\begin{figure}
\includegraphics[width=8cm]{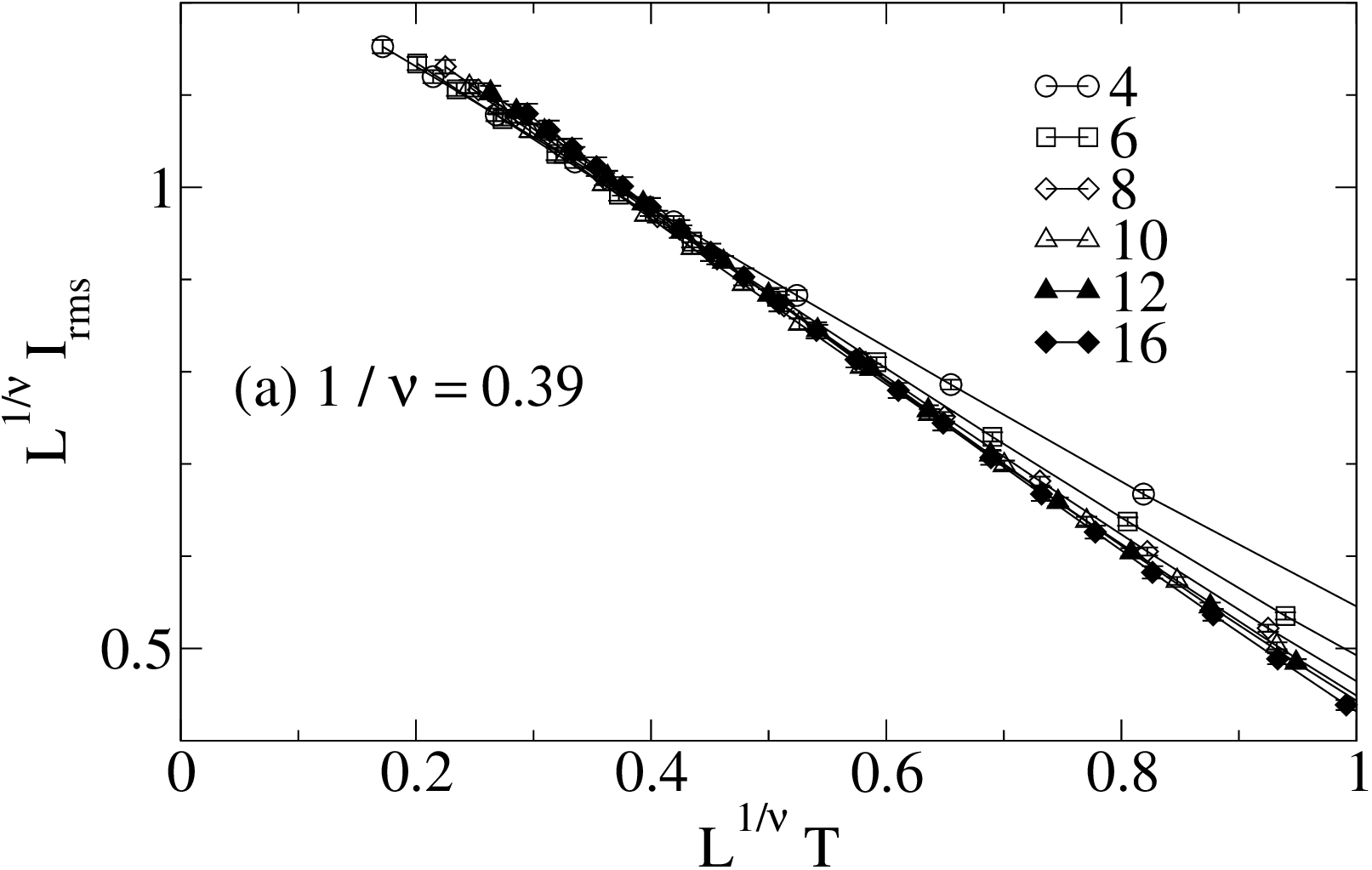}
\includegraphics[width=8cm]{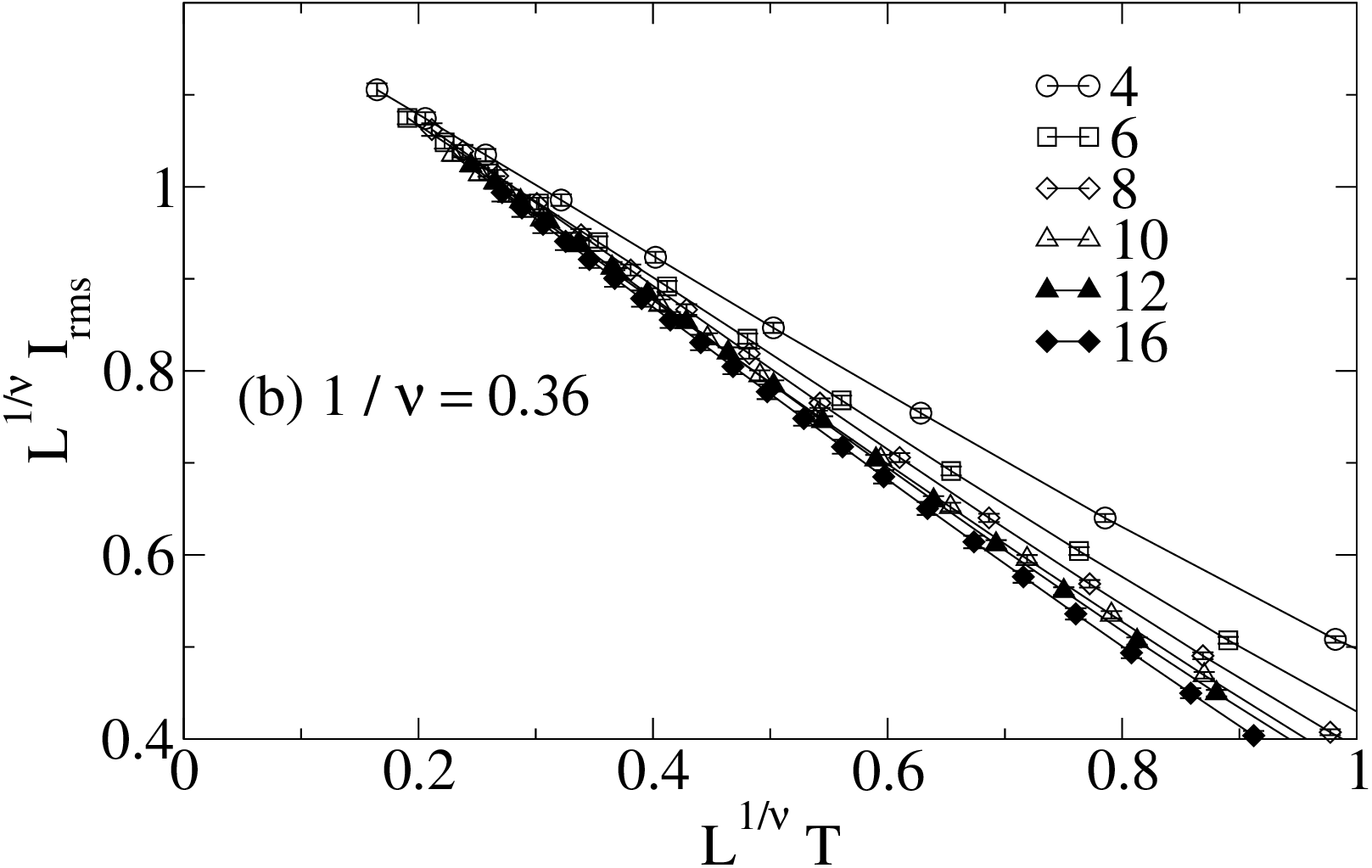}
\caption{Finite size scaling plot of the MC data in Fig.\
\ref{fig.IrmsT} using Eq.\ (\ref{Iscale}) with $T_c=0$.  In (a)
$1/\nu=0.39$, which gives a data collapse except for the lowest
temperatures and smallest system sizes.  In (b) $1/\nu=0.36$, which
gives a better data collapse at the smallest temperatures.}
\label{fig.Iscale}
\end{figure}

To scale MC data for $I_{\rm rms}$ assuming that $T_c=0$, Eq.\
(\ref{Iscale}) shows that a plot of $L^{1/\nu} I_{\rm rms}$ vs.\
$L^{1/\nu}T$ for different $T,L$ should collapse onto a system size
independent curve, for the correct value of $1/\nu$.  Scaling plots
are shown in Fig.\ \ref{fig.Iscale} for the choices (a) $1/\nu =0.39$
and (b) $1/\nu =0.36$.  In (a) we see that $1/\nu =0.39$ gives the
best overall data collapse, but the fit is somewhat poor for the
smallest values of $L^{1/\nu}T$, which for $T_c=0$ is expected to be
the most significant part of the data collapse.  We therefore drop
data points for large $L^{1/\nu}T$ and only fit the value of $\nu$ to
the data points with smallest $L^{1/\nu}T$.  Fitting the data points
at the two lowest temperatures for sizes $L=10,12,16$ to a straight
line, and minimizing the RMS fit error, gives $\nu=0.36 \pm 0.03$.
The error bar is obtained by the bootstrap method \cite{bootstrap}.
The resulting data collapse is shown in Fig.\ \ref{fig.Iscale} (b),
and we see that, for small $L^{1/\nu}T$, a better fit is obtained than
in (a).  The estimate $\nu=0.36 \pm 0.03$ is insensitive within the
error bar to varying the precise details of the fit, such as including
the three lowest temperatures for each $L$.

\section{Results for dynamic quantities}
\label{sec.dynamic}

The dynamics simulations contain only local MC moves, and the averages
computed in the dynamics runs correspond to time integration of the
observables.  The statics simulation contains both local MC moves and
nonlocal temperature exchange moves that take big steps in
configuration space, which corresponds to ensemble averaging.  If
ergodicity is broken these different averages may disagree
\cite{holme02}.  To investigate this issue we plot the RMS current
calculated in a dynamics run in Fig.\ \ref{fig.IrmsT}, together with
the corresponding result from the static simulation.  The figure shows
that the results from static and dynamic runs coincide within error
bars.  

We will make use of the following input from the statics simulations.
The static scaling results gives the value of the correlation
length exponent, which is needed also in the dynamic scaling analysis.
We will assume the value $1/\nu=0.39$, but make occasional comparison
with the corresponding results for $1/\nu=0.36$.  Furthermore, the
static scaling results also give an indication of the range of system
sizes and temperatures where we can expect dynamic scaling of the data
to hold.  From Fig.\ \ref{fig.Iscale} we estimate that dynamic scaling
can be expected when $L^{1/\nu}T$ is smaller than about 0.8.

\begin{figure}
\includegraphics[width=8cm]{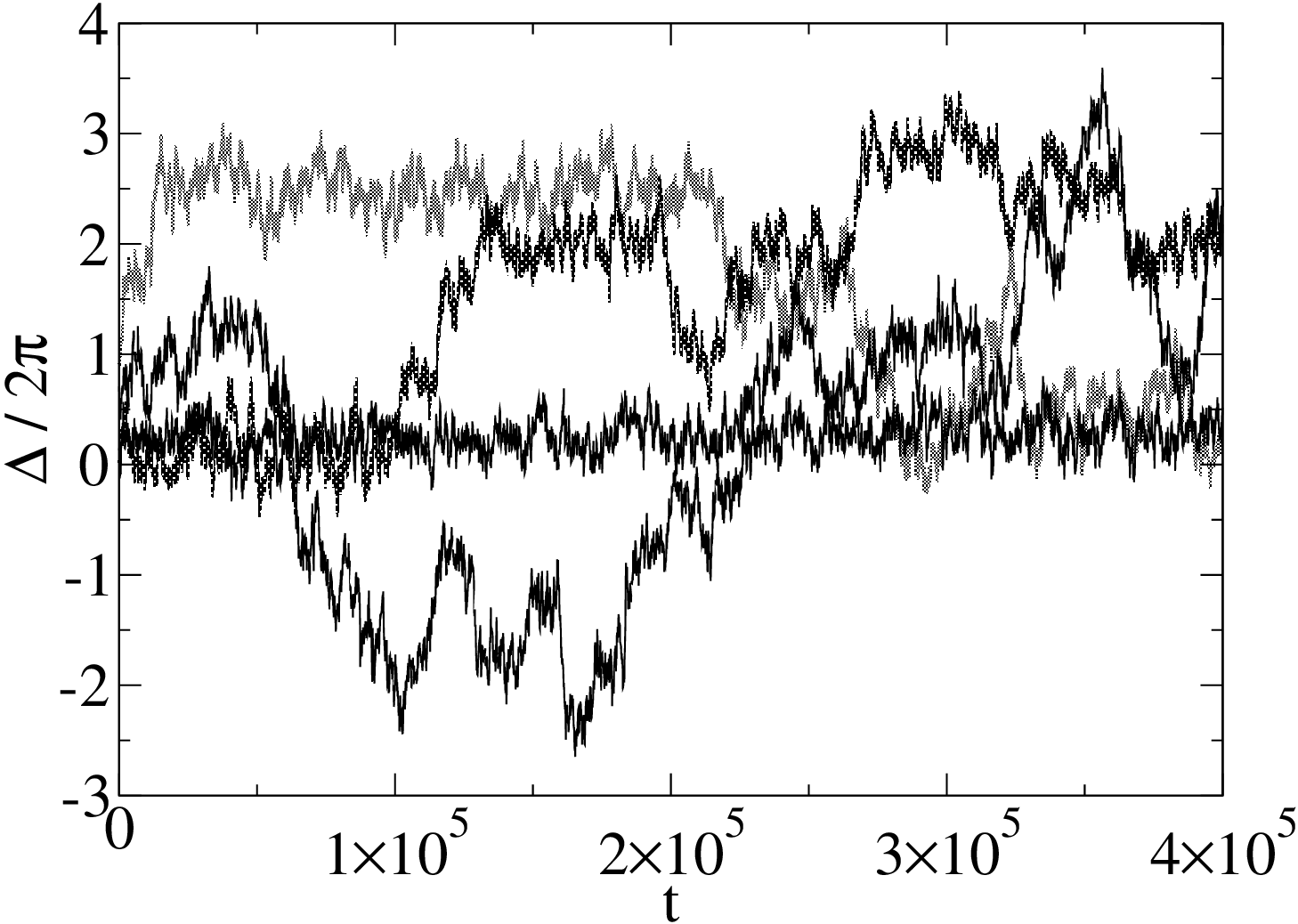}
\caption{Examples from our simulation of the time evolution of MC data
for the phase twist variable for a few different disorder
realizations.  The characteristic time scale for big fluctuations in
the twist variable can be $>10^5$ sweeps.  We identify such
fluctuations as phase slips corresponding to vortex motion
across the sample.  
}
\label{fig.slip}
\end{figure}

We expect that the longest emergent time scale comes from phase slip
events corresponding to vortex motion across the sample.  An
illustration of these long correlation times is seen in Fig.\
\ref{fig.slip}, which shows phase slip events in a dynamic simulation
where fluctuations in the boundary conditions are included, for a few
different realizations of the quenched disorder.  Phase slip events
are often as far as about $10^5$ sweeps apart.

\begin{figure}
\includegraphics[width=8cm]{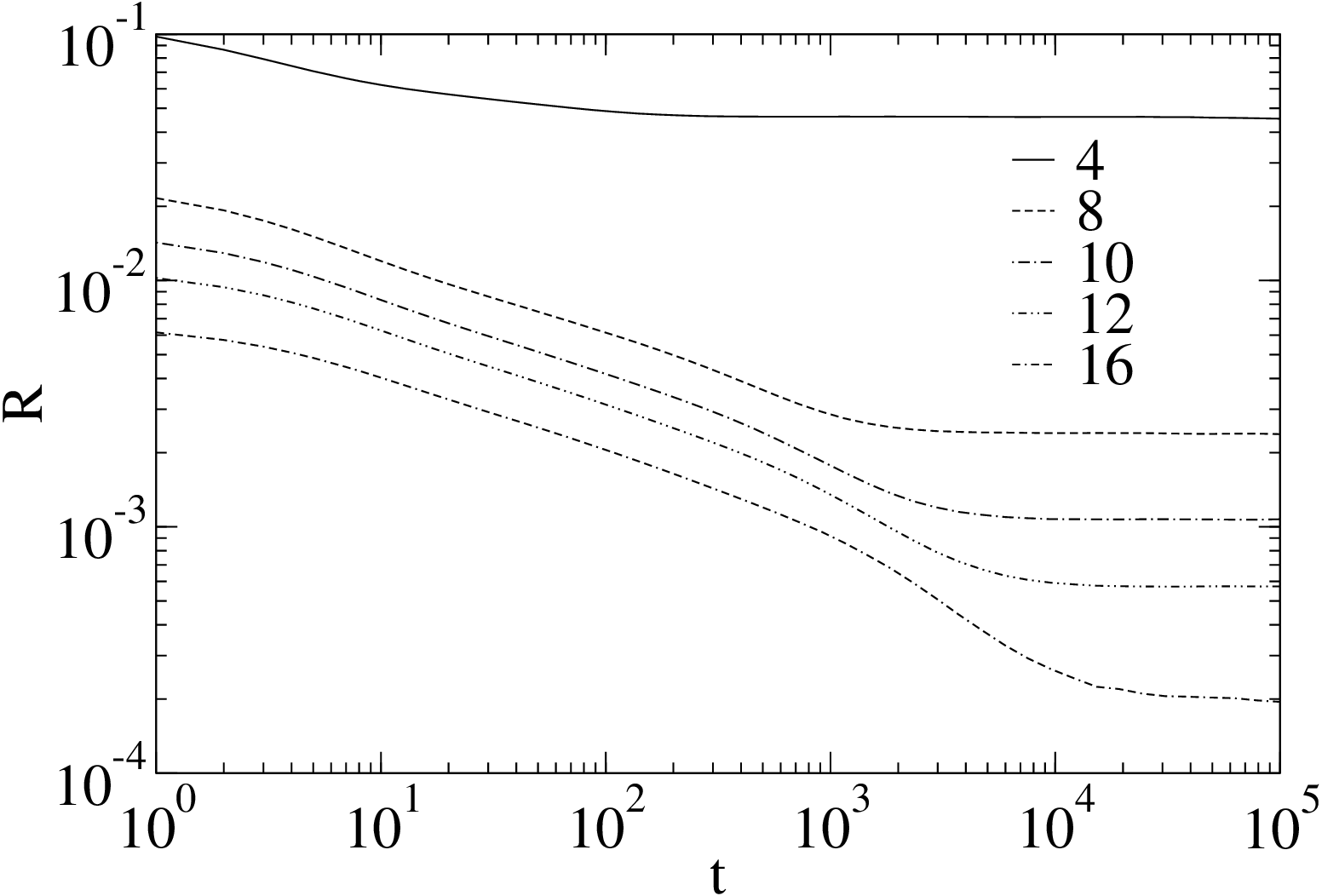}
\caption{MC data for the resistance function $R(t,T,L)$ vs.\ time $t$
for different system sizes for $A=L^{1/0.39}T=0.8$.  The resistance
curves have different characteristic regimes, one for $t > \tau$
where $R$ becomes independent of $t$, and one power-law regime for
shorter times.}
\label{fig.Rt}
\end{figure}

Figure \ref{fig.Rt} shows the time dependent resistance function given
by Eq.\ (\ref{R}) for $L^{0.39}T=0.8$.  The data curves show
unexpectedly complicated behavior, with a power-law regime at short
times, and a time-independent limit at long times, with various
crossover times in between.  We will now analyze the scaling behavior
of these different time scales.

\begin{figure}
\includegraphics[width=8cm]{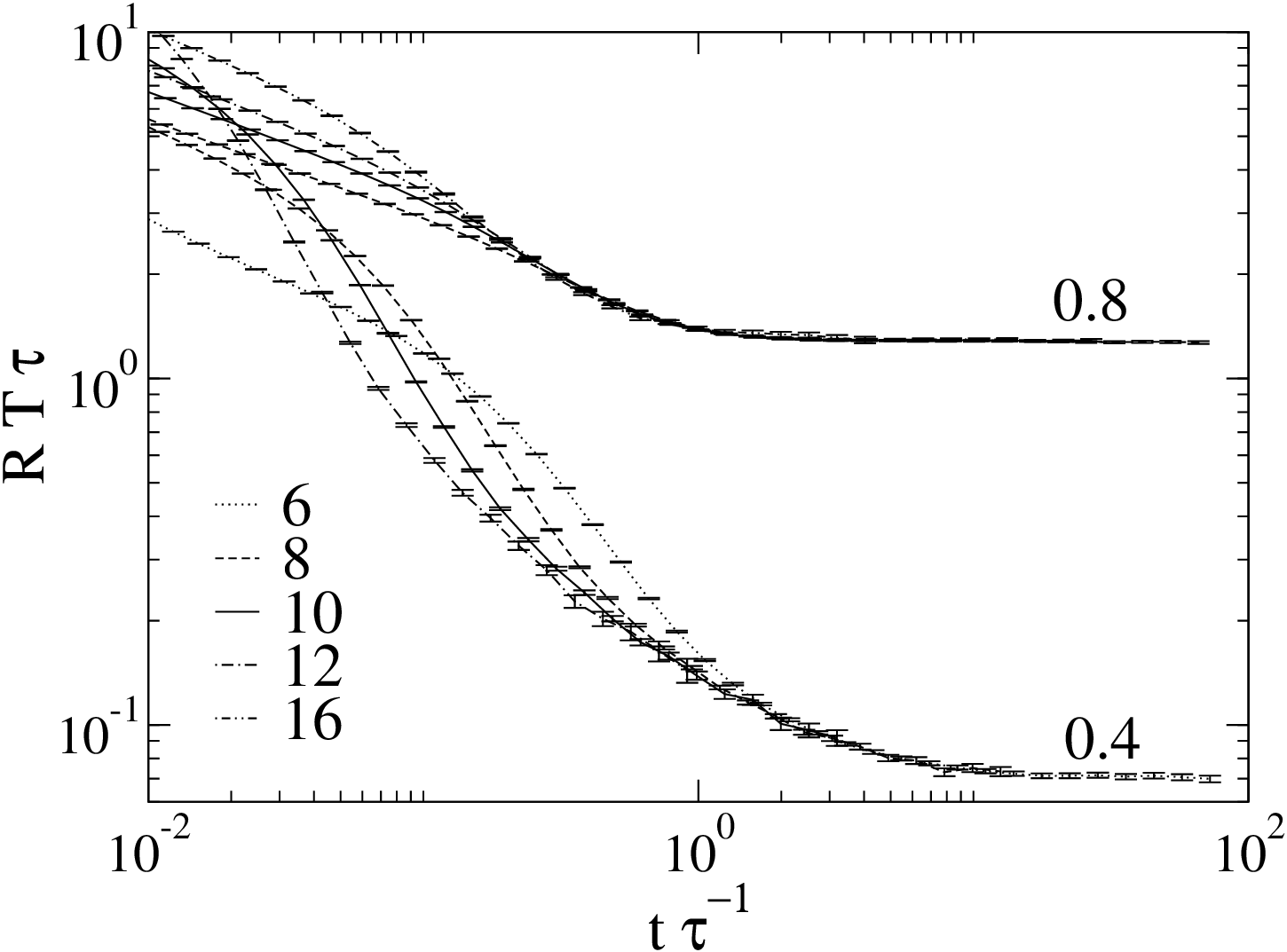}
\caption{MC data for $RT\tau$ vs.\ $t/\tau$ for different temperatures
and system sizes with $A=L^{0.39}T=0.4$ and $A=0.8$.  For each value
of $T,L$ the correlation time $\tau$ has been tuned until an optimal
data collapse is obtained in the large $t$ limit.}
\label{fig.Rscale}
\end{figure}

We first consider the long time limit, $t \gg \tau$, of the data curves
in Fig.\ \ref{fig.Rt}, which gives the resistance $R$.  According to
Eq.\ (\ref{Rscale}), the correct value of $\tau$ for each system size
and temperature, such that $L^{1/\nu}T=A$ is a constant, should make
data curves for $RT\tau$ plotted vs.\ $t/\tau$ collapse onto a common
curve for $t \gg \tau$.  Thus we adjust $\tau$ for each data curve to
the value that gives the best collapse.  The result is shown in Fig.\
\ref{fig.Rscale}.  We obtain a data collapse to a common curve at
large $t$, for each set of data curves belonging to the same value for
$A=L^{1/\nu}T$.  At shorter times the collapse breaks down, indicating
a different short time behavior.  By this procedure we determine an
estimate of the correlation time $\tau(T,L)$, up to an unknown
multiplicative constant that depends on $A=L^{1/\nu}T$.  The scaling
result shown in this figure demonstrates that, up to the level of
statistical accuracy of the simulation data, the resistance obeys a
dynamic scaling relation which assumes $T_c=0$.  This result is
independent of any assumption about the functional form of
$\tau(L,T)$.

\begin{figure}
\includegraphics[width=8cm]{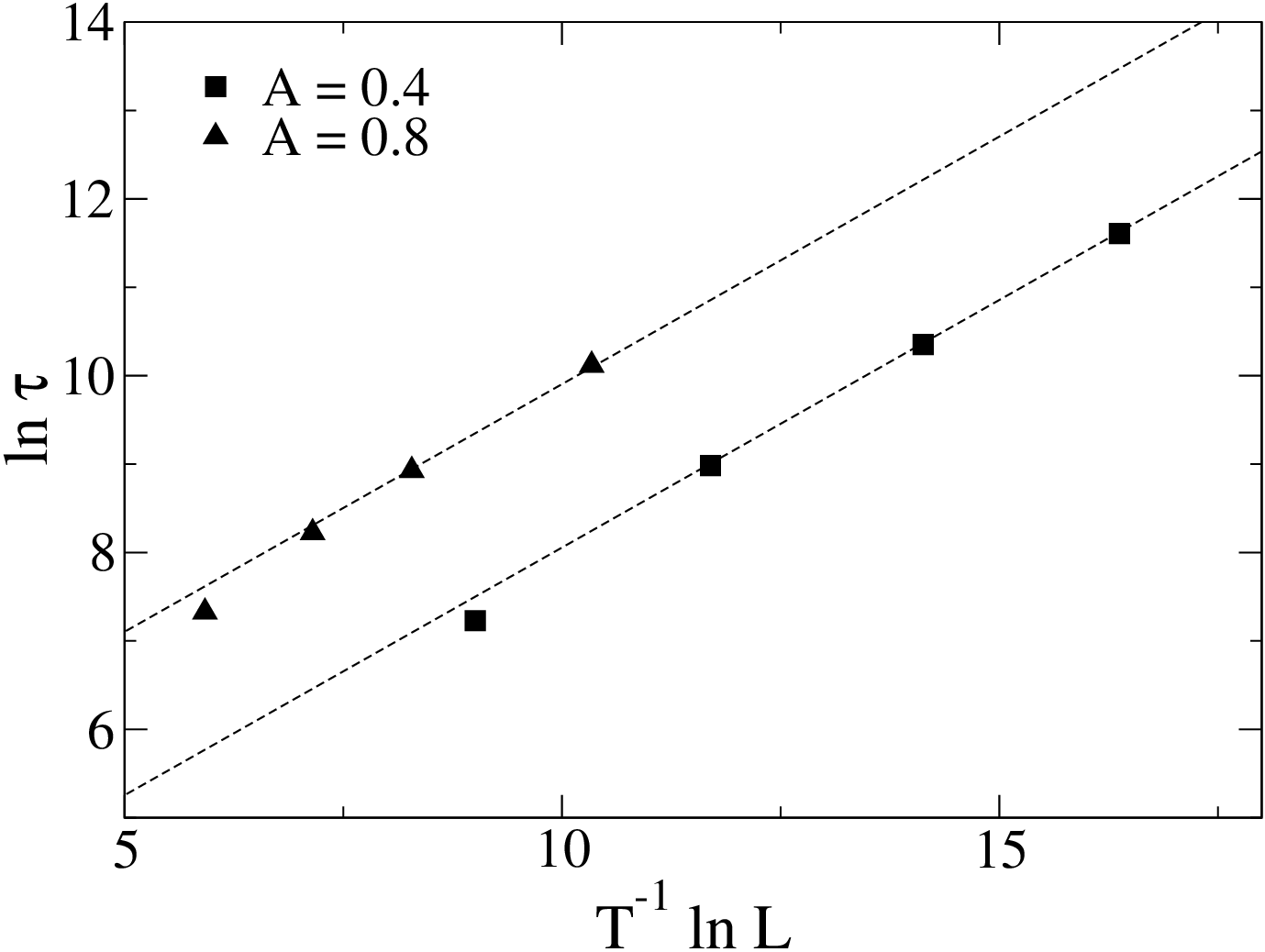}
\caption{MC data for the correlation time $\tau$, determined in Fig.\
\ref{fig.Rscale}, vs.\ $(\ln L)/T$ for different temperatures and system
sizes with $A=L^{0.39}T=0.4$ and $A=0.8$.  The dashed lines correspond
to $\tau \sim \exp K(\ln L)/T$, with $K=0.56$ independent of $T,L$.}
\label{fig.tau}
\end{figure}

To proceed further, we need to fit the values for $\tau$ and we then
need an assumption of a functional form for $\tau$.  All the
functional forms $\tau \sim \exp K(\ln L)/T$, $\tau \sim \exp K
L^\psi/T$, and $\tau \sim L^z$ approximately fit the data, and our
present data is therefore not sufficient to clearly distinguish
between these functional forms.  Larger system sizes and lower
temperatures are necessary in order to discriminate between them.
The form $\tau \sim \exp K(\ln L)/T$ gives a good fit to the data,
with $K\approx 0.56$ independent of $T$ and $L$, shown in Fig.\
\ref{fig.tau}.  However, this should only be considered as one
possible parameterization of the correlation time.

Next we consider the scaling behavior of $R(t,T,L)$ for intermediate
times.  Surprisingly the scaling found for long times in Fig.\
\ref{fig.Rscale} is not obeyed for intermediate times where the data
collapse breaks up.  We therefore attempt another scaling relation to
apply in this regime.  According to Eq.\ (\ref{Rscale}), we again set
$L^{1/\nu}T=A$ to a constant value, and plot $RT\tau$ vs.\ $t/\tau$.
We can then obtain an approximate data collapse in the intermediate
power-law regime in the figure for $\tau \sim L^z$ with $z= 2.7 \pm
0.2$, where the error bar is estimated from the interval outside of
which a significantly worse fit is obtained.  We can also fit the same
data with other functional forms for $\tau$, but we get the best fit
with $\tau \sim L^z$.  These results show that the resistance has two
separate dynamic scaling regimes, characterized by two separate
diverging correlation times.  It is natural to associate the longest
of these correlation times with the phase slip processes that make the
resistance finite at any finite temperature, since the correlation
time roughly matches the time between phase slips in Fig.\
\ref{fig.slip}.  We will now study these different dynamic scaling
regimes in more detail.

\begin{figure}
\includegraphics[width=8cm]{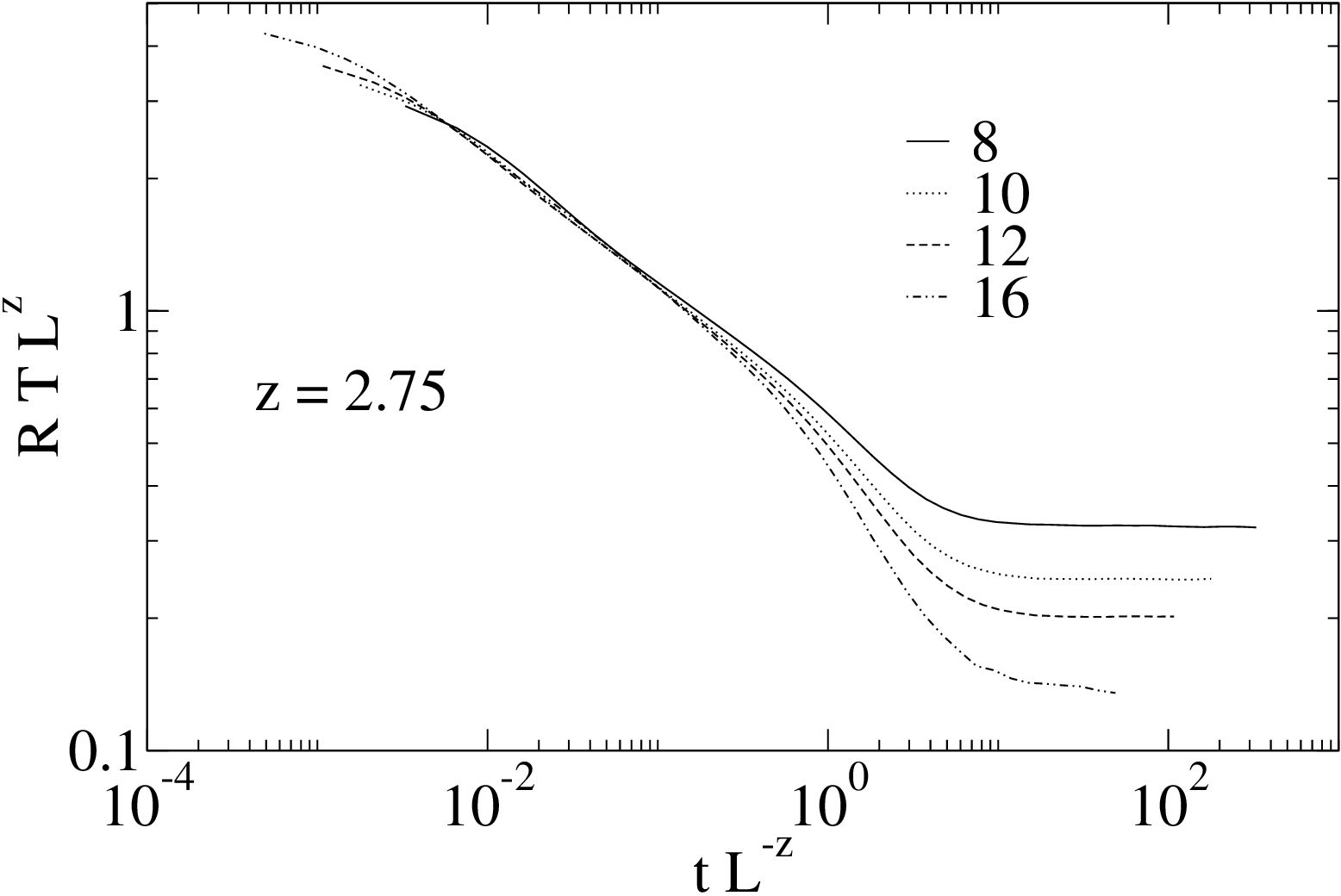}
\caption{Finite size scaling collapse of MC data for the time
dependent resistance function $R(t,T,L)$ with system sizes $L$ and
temperatures $T$ related by $A=L^{1/\nu}T=0.8$.  The data collapse is
obtained for $\tau \sim L^z$ with $z=2.75$.  Data for other values of
$A$ scale with nearly the same value of $z$.  }
\label{fig.Rscale.middle}
\end{figure}

\begin{figure}
\includegraphics[width=8cm]{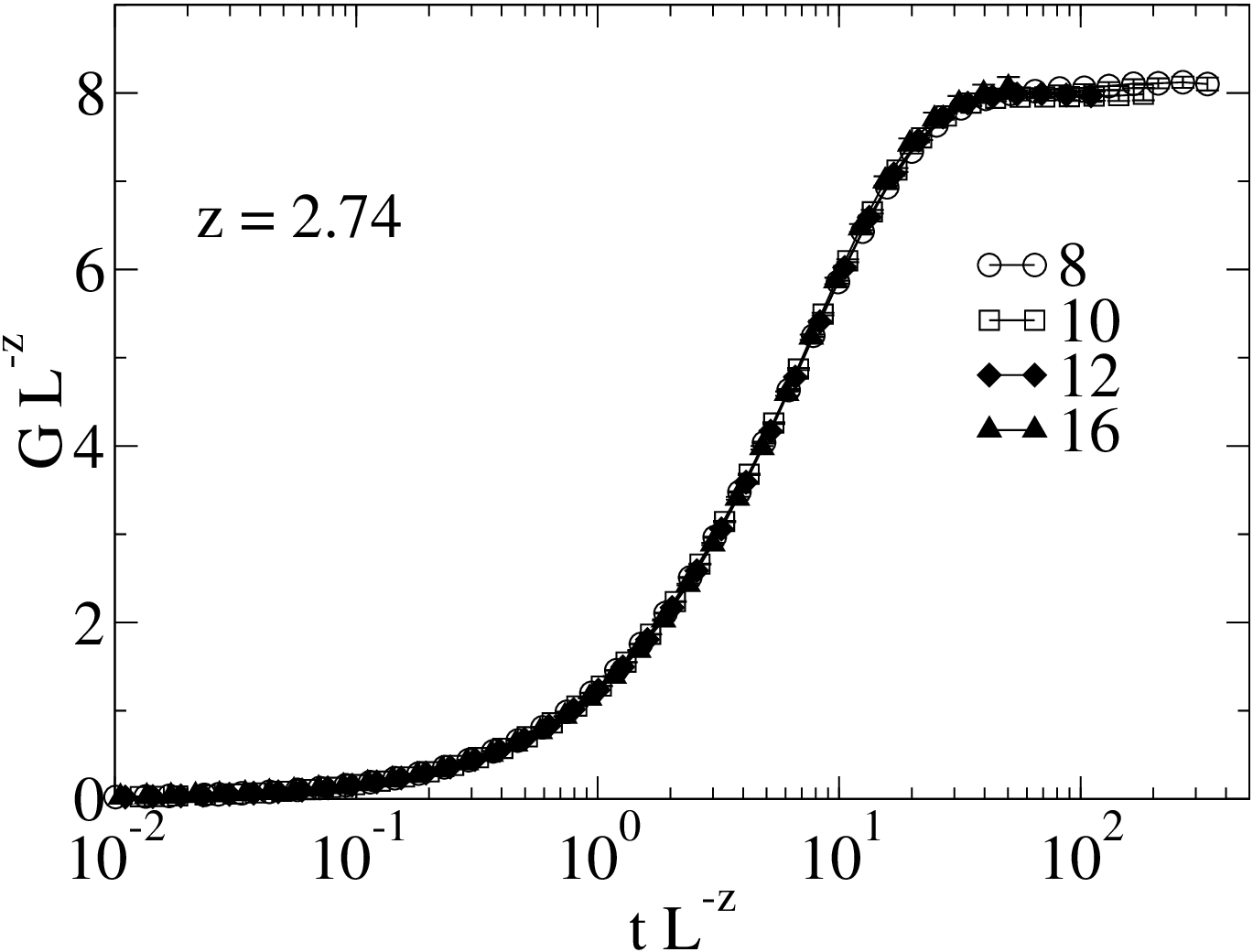}
\caption{Finite size scaling data collapse of MC data for the dynamic
correlation function $G(t,T,L)$ defined in Eq.\ (\ref{G}).  }
\label{fig.Gscale}
\end{figure}

To reach further insight into the scaling properties at intermediate
times we consider the order parameter autocorrelation function
$G(t,T,L)$, defined in Eq.\ (\ref{G}).  Figure \ref{fig.Gscale} shows
MC data for $G(t,T,L)$ scaled according to Eq.\ (\ref{Gscale}) with
$\tau \sim L^z$.  The data in the figure has $L^{1/\nu}T=0.8$, and the
data collapse is obtained for $z=2.74$.  The data in the figure is for
the case of fluctuating phase twists at the boundaries.  We also
calculated the same quantity with periodic boundary conditions, and
obtained an equally good fit with the same exponent.  Approximately
the same exponent also gives a collapse for other values of $A$, but a
slight drift in $z$ is noticeable, just as the data collapse in Fig.\
\ref{fig.Iscale} for the RMS current is somewhat dependent on $A$.
Including this uncertainty we estimate $z=2.7 \pm 0.2$.  We conclude
that the dynamics measured by $G$ is independent of the phase twist
fluctuations.  The dynamic scaling relation $\tau \sim L^{2.7}$,
instead comes from local dynamic fluctuations other than the global
phase slips.  This means that we can interpret the intermediate
resistance scaling regime in Fig.\ \ref{fig.Rscale} as coming from
local fluctuations rather than global phase slips.

\section{Discussion and conclusions}
\label{sec.discussion}

This paper considers finite size scaling properties of Monte Carlo
data for the glass transition in the two-dimensional gauge glass
model.  Equilibration of our simulations at low temperatures is
achieved by a temperature exchange simulation method.  Up to the level
of statistical uncertainty, system sizes and temperatures reached in
the simulations, our finite size scaling results for both static and
dynamic properties show scaling properties that are clearly consistent
with a glass transition occurring only at zero temperature, and thus
indicates the absence of a superconducting glass state at any finite
temperature in 2D.

Our results for static scaling properties reproduces, and adds some
details to, some of the results of Refs.\
\onlinecite{katzgraber02,katzgraber03}, in addition to providing a
useful starting point for our dynamics simulations.  From the scaling
of the RMS current we find the correlation length exponent given by
$1/\nu \approx 0.39$, obtained by optimizing the quality of a finite
size scaling data collapse over a finite temperature interval close to
$T=0$.  This value agrees with that of Refs.\
\onlinecite{katzgraber02,katzgraber03}.  If we instead optimize the
fit at the lowest temperatures where we have data, where scaling is
also expected to work best, we obtain $1/\nu = 0.36 \pm 0.03$.  It is
interesting to note that this is in very good agreement with the
result for the ground state domain wall exponent $\theta=-1/\nu$ in
Ref.\ \onlinecite{akino02}.  To obtain an even more accurate estimate
of $\nu$, data for lower temperatures and bigger system sizes is
necessary.  To converge such data would require a considerable
increase in simulation efforts.

For the dynamic properties, we consider the linear resistance
computed from the diffusion of the phase slip across the boundary of
the system.  Since the correlation times grow very long at low
temperature, it is not possible to converge the simulation at as low
temperature as for the static data where the temperature exchange
algorithm helps considerably, which makes it more difficult to study
dynamic scaling than static scaling.  A further complication is that
our data for the linear resistance shows more than one characteristic
time scale.  We identify the longest correlation time as the
characteristic time scale for phase slip fluctuations across the
sample.  Unfortunately it is not possible to determine the correlation
time uniquely from the data, since the interval of sizes and
temperatures where we have data is too limited.  All the forms $\tau
\sim \exp KL^\psi/T$, $\tau \sim \exp K \ln L /T$, and $\tau \sim L^z$
can be made to approximately fit the data.  To discriminate between
the possible forms, data at lower $T$ and larger $L$ is needed, which
is beyond the present calculation.  However, a logarithmically growing
phase slip barrier, with $\tau \sim \exp K \ln L /T$, gives a good fit
to the data with $K\approx 0.56$.  A qualitatively similar result was
obtained in a zero temperature calculation of the phase slip barrier
energy in Ref.\ \onlinecite{holme02}.  Since $R \sim \exp -KL^\psi/T$,
of which the logarithmic barrier is a special case with $\psi=0$, a
superconducting glass with zero resistance is obtained only for $T=0$.
In other words the dynamic glass transition occurs at zero
temperature.

Surprisingly, data for the time dependent resistance function
$R(t,T,L)$ does not scale with the longest correlation time for
intermediate times.  Instead, in this regime scaling is obtained by
assuming a new correlation time of the form $\tau \sim L^z$ with
$z= 2.7 \pm 0.2$.  Hence the scaling behavior of our data implies two
different diverging correlation times with quite different properties
at the zero temperature glass transition.  A further indication of the
smaller correlation time is given by the scaling of the order
parameter autocorrelation function, which scales over the entire time
interval where we have data, including the longest times, quite in
contrast to the resistance.  The interpretation of this shorter
correlation time is that it gives the characteristic time scale for
fluctuations other than the global phase slip fluctuations that lead
to a finite resistance.  Similar results showing different dynamics
for local fluctuations and phase twist fluctuations have been obtained
previously at the finite temperature transition in XY models without
disorder \cite{jensen}.  
We also note that the scaling relation
$\tau \sim L^z$ can be written $\tau \sim \exp z \ln L \sim \exp z
L^0 \sim \exp C/T^{1+\psi\nu}$, which means that
$\psi=\theta=-1/\nu$.  Hence the relation $\tau \sim L^z$ suggests
that the static and dynamic free energy barrier exponents are the same
for the local fluctuations of the order parameter.

The two different diverging correlation times suggested here may
explain why signatures of a glass phase with zero resistance at finite
temperatures can easily be obtained in simulations.  For the shorter
correlation time we have $\psi=\theta=-1/\nu$, which gives $R(T) \sim
\exp -C/T^{1+\psi \nu}=const$, for the resistance at small
temperatures, i.e.\, the resistance stays at the $T=0$ value $R=0$ for
small $T$.  This shows that in a regime of intermediate time scales,
zero resistance can be obtained at finite temperature.  However, at
much longer time scales that may be hard to reach in a simulation,
phase slip fluctuation set in, and qualitatively different behavior
follows with a superconducting glass transition only at zero
temperature.

In summary, from finite size scaling of a time dependent resistance
function and an order parameter autocorrelation function, we
constructed two different diverging correlation times from the $T_c=0$
scaling behavior of Monte Carlo data.  For further work, it would be
of interest to further clarify the precise details of dynamic scaling
at the zero temperature glass transition, and to include quantum
fluctuations in the model.  It would furthermore be interesting to
search for the proposed novel dynamic scaling behavior in experiments.

We acknowledge very helpful discussions with Petter Minnhagen, Helmut
Katzgraber, Peter Olsson, Stephen Teitel, and Peter Young.  This work
was supported by the Swedish Research Council, PDC, NSC, and the
G{\"o}ran Gustafsson foundation.

\bibliography{gaugeglass}
\end{document}